\documentclass{aa} 
\usepackage{newtxtext,newtxmath}

\usepackage[T1]{fontenc}
\usepackage[normalem]{ulem}

\usepackage{graphicx}   
\usepackage{amsmath}    
\usepackage{amssymb}    
\usepackage{color}
\usepackage{booktabs}
\usepackage{xspace}
\usepackage{bm}
\usepackage{verbatim}
\usepackage{float}
\usepackage{tensor}
\usepackage{xcolor}
\usepackage{siunitx}
\usepackage[switch]{lineno}
\usepackage[normalem]{ulem}
\newcommand{\orcid}[1]{\href{https://orcid.org/#1}{\includegraphics[width=8pt]{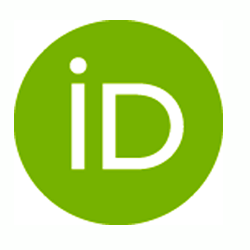}}}
\makeatletter

\renewcommand*{\@fnsymbol}[1]{\ensuremath{\ifcase#1\or \href{https://relativist1.github.io/}{\includegraphics[height=2.4ex]{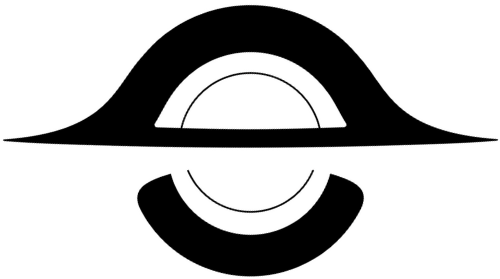}}\or \includegraphics[height=3.8ex]{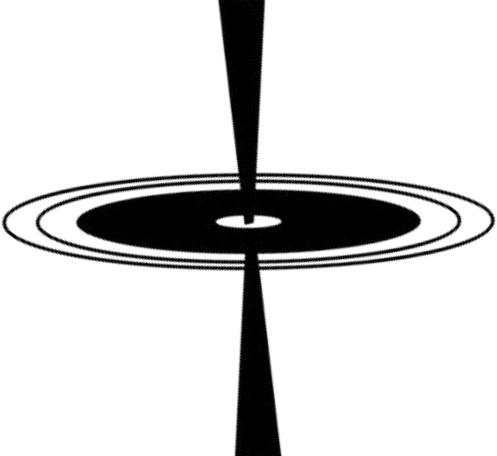}\or \ddagger\or
   \mathsection\or \mathparagraph\or \|\or **\or \dagger\dagger
   \or \ddagger\ddagger \else\@ctrerr\fi}}
\makeatletter
\usepackage[colorlinks]{hyperref}
\hypersetup{
    colorlinks=true,
    linkbordercolor={white},
    linkcolor={blue},
    citecolor={blue}
} 
\graphicspath{{./}{figures/}}

\newcommand{\be}{\begin{equation}}
\newcommand{\ee}{\end{equation}}

\def\muas{\mu\rm{as}}
\def\m87{M\,87*}

\begin{document}

\title{Semi-analytic studies of the accretion disk and magnetic field geometry in M\,87*}
\titlerunning{Semi-analytic studies of the accretion and magnetic fields in \m87}
\authorrunning{Saurabh et. al.}
\author{Saurabh\orcid{0000-0001-7156-4848}\inst{1}\fnmsep\thanks{\email{saurabh@mpifr-bonn.mpg.de} \\ 
Member of the International Max Planck Research School (IMPRS) for Astronomy and Astrophysics at the Universities of Bonn and Cologne.}  
\and 
Maciek Wielgus\orcid{0000-0002-8635-4242}\inst{2}\fnmsep\thanks{\email{maciek@wielgus.info}} 
\and 
Arman Tursunov\orcid{0000-0001-5845-5487}\inst{1,3}  
\and 
Andrei P. Lobanov\orcid{0000-0003-1622-1484}\inst{1} 
\and
Razieh Emami\orcid{0000-0002-2791-5011}\inst{4}
}
\institute{Max-Planck-Institut f\"ur Radioastronomie, Auf dem H\"ugel 69, D-53121 Bonn, Germany 
\and 
Instituto de Astrofísica de Andalucía-CSIC, Glorieta de la Astronomía s/n, E-18008 Granada, Spain
\and
Research Centre for Theoretical Physics and Astrophysics, Institute of Physics, \\Silesian University in Opava, CZ-74601 Opava, Czech Republic 
\and
Center for Astrophysics $\vert$ Harvard \& Smithsonian, 60 Garden Street, Cambridge, MA 02138, USA
\date{}
}
\abstract
{
Magnetic fields play a pivotal role in the dynamics of black hole accretion flows and in the formation of relativistic jets. Observations by the Event Horizon Telescope (EHT) provided unprecedented insights into accretion structures near black holes. Interpreting these observations requires a theoretical framework that links polarized emission to the underlying system properties and magnetic field geometries.
}
{ 
We investigated how the system properties, in particular, the magnetic field geometry in the region of the event horizon scale, affect the structure of the observable synchrotron emission in \m87. Specifically, we characterized the sensitivity of observables used by the EHT to black hole spin, plasma dynamics, accretion disk thickness, and magnetic field geometry.
}
{
We adopted a semi-analytic radiatively inefficient accretion flow model in Kerr spacetime. We varied the magnetic field geometry, black hole spin, accretion disk dynamics, and geometric thickness of the disk. We performed general relativistic ray-tracing with a full polarized radiative transfer to obtain synthetic images of \m87. We extracted EHT observables, such as disk diameter, asymmetry, and polarimetric metrics from synthetic models. We also considered a number of general relativistic magnetohydrodynamics simulations and compared them with the semi-analytical models.
}
{
The effect of the disk thickness on the observables is limited. On the other hand, magnetic configurations dominated by the toroidal and poloidal fields can be distinguished reliably. The flow dynamics, in particular, radial inflow, also significantly affects the EHT observables. 
}
{
The \m87 system is most consistent with a flow dominated by the poloidal magnetic field with partially radial inflow. While the spin remain elusive, moderate or high positive values are preferred.
}
\bigskip

\maketitle

\section{Introduction}
It is unclear how magnetic fields govern the behavior of matter near black holes (BHs), as is the role they play in the formation of powerful jets and accretion flows in these extreme environments. This question remains central to our understanding of BH astrophysics. The interplay between magnetic fields and accretion flows significantly affects the immediate vicinity of BHs and extends to broader implications for galaxy dynamics and the evolution of supermassive BHs in active galactic nuclei \citep[AGN;][]{1977MNRAS.179..433B, Fabian2012}. 
At the event horizon scale, magnetic fields are thought to be crucial for angular momentum transport and energy dissipation within accretion flows \citep{1991ApJ...376..214B}, thereby affecting the accretion rate and playing a key role in launching and collimating relativistic jets \citep{2005AJ....130.2473K, 2008ApJ...678.1180B}. These processes can be modeled with the equations of magnetohydrodynamics \citep[MHD; e.g.,][]{gammie2003, 2012MNRAS.423.3083M}, with the structure and evolution of magnetic fields being key to deciphering the observed properties of BH systems, including their polarized emission. 

It was not until the recent radio observations of the centers of galaxies Messier 87 (\m87) and the Milky Way, Sagittarius A* (Sgr\,A*), by the Event Horizon Telescope \citep[EHT;][]{2019ApJ...875L...1E,2019ApJ...875L...4E, SgrA_P1_2022,SgrA_P3_2022} and the near-infrared (NIR) observations by the GRAVITY instrument \citep{2018A&A...615L..15G} that new avenues were opened to probe horizon-scale properties of BHs. EHT observations of \m87 and Sgr\,A* revealed compact bright ring-like emission structures with substantial central brightness depression, providing compelling evidence for the existence of supermassive compact objects in their cores \citep{2019ApJ...875L...1E,SgrA_P1_2022}. 
Subsequent analyses of the EHT observations yielded further insights into the physics of these systems, particularly through the study of polarized emission structures. They constrained the magnetic field geometry in the core region \citep{2021ApJ...910L..12E,2021ApJ...910L..13E,SgrA_P7_2024,SgrA_P8_2024}. These observations revealed dynamically important magnetic fields that are expected to play a crucial role in launching and collimating relativistic jets.

In order to explain the observations and model the behavior of magnetized plasma in the strong gravitational field near BHs, general relativistic magnetohydrodynamics (GRMHD) simulations are employed \citep{gammie2003, 2004ApJ...611..977M}. These simulations have become an indispensable tool for investigating the intricate interplay between magnetic fields \citep{2012MNRAS.423.3083M, 2017ApJ...837..180G,2022ApJ...924L..32R}, accretion flows \citep{ 2019ApJS..243...26P,2022MNRAS.511.3795N,2022MNRAS.516.5092M}, and spacetime geometry \citep{2018NatAs...2..585M, 2023arXiv231020040C}, and their combined effects \citep{2019ApJ...875L...1E, 2020MNRAS.499..362C, 2022MNRAS.513.4267N}. 

While GRMHD simulations offer the most physically self-consistent framework for studying BH accretion, the inference based on these models remains contentious \citep[e.g.,][]{Gralla2021,Wielgus2021}. Despite its successes as an effective model, the set of equations solved by GRMHD is inadequate to describe collisionless plasma, assuming highly uncertain initial conditions. Furthermore, most of the results are obtained within the ideal MHD framework, while the important role of nonideal MHD effects such as magnetic reconnection in turbulent accretion flows near the event horizon is recognized \citep[e.g.,][]{2022ApJ...924L..32R}. Global stationary semi-analytic approaches offer a complementary perspective by enabling parametric studies of accretion flows and magnetic field geometries. They allow us to investigate the effect of the prescribed disk properties and magnetic field geometry on the observable properties, including polarization of the emitted radiation \citep{1995ApJ...452..710N, 2018ApJ...863..148P}. This approach constitutes a bridge between toy models that are useful for building intuition \citep[e.g.,][]{Narayan2021} and computationally exhaustive GRMHD, where the parameters of the accretion disk and magnetic field follow the evolution of the MHD equations and cannot be modified explicitly.

A low-luminosity AGN system of \m87 falls into the category of radiatively inefficient accretion flows \citep[RIAFs;][]{2003ApJ...598..301Y} with a geometrically thick and optically thin accretion disk. Unlike optically thick and geometrically thin disks \citep{Shakura1973}, which efficiently cool down by blackbody radiation, RIAFs are characterized by low number densities and an extremely high temperature of ions that is decoupled from that of the radiating electrons. Thus, radiative cooling is inefficient, and excess energy is removed from the system through advection. A stationary semi-analytic RIAF model framework for \m87 was proposed \citep[e.g.,][]{2009ApJ...697.1164B, Vincent2021}. We systematically study the effect of the magnetic field configuration, BH spin, as well as geometric thickness and dynamics of the accretion disk on the observational appearance of RIAF models. We compare the RIAF results with GRMHD simulations, as well as with the characteristic observables extracted from the EHT data, including polarimetric quantities.   

\begin{figure*}
    \centering
    \includegraphics[width=0.9\textwidth]{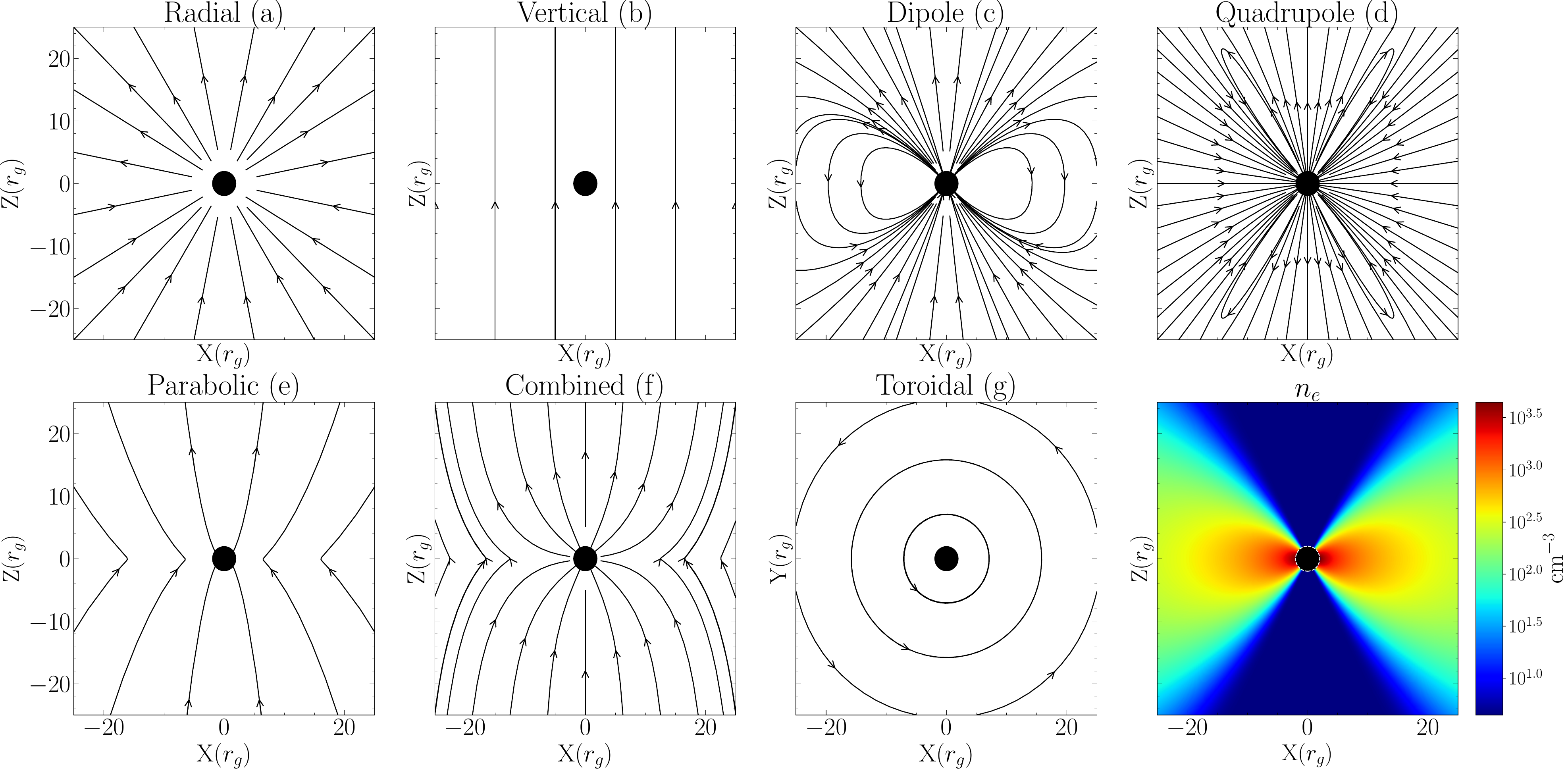}
    \caption{Panels a-g: Schematic representation of the magnetic field configurations discussed in Sect.~\ref{sec:model}. Last panel: Example number density map for an RIAF model with a default disk thickness parameter $H=0.5$.  } 
    \label{fig:bfield_schematic}
\end{figure*}
\section{Model images of a black hole}  \label{sec:model}


We set up a stationary axisymmetric RIAF model in Kerr spacetime and subsequently performed numerical general relativistic ray-tracing using the open-source code \texttt{ipole}\footnote{\url{https://github.com/AFD-Illinois/ipole}} \citep{2018MNRAS.475...43M}. We assumed that the observed emission originates near the equatorial plane in the accretion flow, and we thus neglected potential contributions from the jet base. While the jet plays a major role for the source morphology and its energetic output at lower radio frequencies \citep[e.g.,][]{Walker2018}, the compact ring-like morphology of the 230\,GHz EHT images of \m87 was found to be more consistent with the emission dominated by the accretion flow \citep[e.g.,][]{2019ApJ...875L...5E, 2021ApJ...910L..13E, 2023ApJ...950...38E}. This is particularly true for the favored strongly magnetized models (see \citet{2019ApJ...875L...5E} and Sect.~\ref{subsec:GRMHD}). While the extended jet in \m87 has been detected at 230\,GHz \citep{Goddi2021}, observations support its subdominant role on compact scales at this frequency. The emission component corresponding to the jet on scales of $\sim\!100$ Schwarzschild radii has been weakly constrained only most recently in the EHT datasets \citep{EHT2025,saurabh2025}. Nonetheless, even if the emission is dominated by the accretion disk, the foreground jet may still act as a Faraday screen, modifying some polarimetric observables \citep{Goddi2021,Wielgus2024}. To capture these effects in the semi-analytic framework, a more complicated two-component disk + jet model would be necessary \citep[see, e.g., preliminary work of][]{Vincent2019}. We provide relevant details of the disk model we used below.

\begin{table} 
    \centering
    \caption{Fiducial accretion flow model parameters}
    \begin{tabular}{c c c}
        \hline
        Parameter & Value & Parameter Description\\
        \hline
        $M_{\rm BH}$ & $6.4\times10^9\rm{M}_\odot$ & Black hole mass \\
        $D_{s}$ &  $16.9 \times 10^6 \rm{pc}$ & Distance to the source\\
        $\theta_i$ & $163^\circ$ & Inclination angle\\
        $\delta$ & 1.5 & $n_{\rm e}$ power law index\\
        $\gamma$ & 0.84 & $T_{\rm e}$ power law index\\
        $H$ & $0.5$ & Geometric thickness\\

        $\kappa_{\rm K}$ & $0.5$ & Keplerian parameter\\
        $\kappa_{\rm ff}$ & $0.5$ & Radial infall parameter\\
        $B_0$ & $5$ G & Magnetic field strength\\
        $n_{\rm e,0}$ & $\approx10^4\,\rm{cm}^{-3}$ & Number density\\
        $T_{\rm e,0}$ & $\approx10^{11}\,\rm{K}$ & Temperature of electrons\\
        $\nu_{\rm obs}$ & $230$\,GHz & Observing frequency\\
        $p_0$ & -100  & Combined field parameter\\
        FOV & $200\,\mu\rm{as}$ & Field of view\\
        $N_X \times N_Y$ & $200 \times 200\,\,\rm{px}$  & Number of pixels\\
        \hline
    \end{tabular}
    \label{table:param}
\end{table}
\subsection{RIAF setup} \label{sec:flow_description}
We adopted a simple semi-analytic RIAF model of a radiatively inefficient accretion flow \citep[e.g.,][]{2016ApJ...820..137B, 2018ApJ...863..148P,2022A&A...668A.185V}. In this framework, we prescribed a stationary model of the accretion flow instead of evolving conservative MHD equations. In particular, we assumed that the number density of electrons $n_{\rm e}$ and electron temperature $T_{\rm e}$ behaves as power laws of the spherical radius,
\begin{align}
n_{\rm e} &= n_{\rm e,0} \left( \frac{r}{r_g} \right)^{-\delta} \exp \left[- \frac{1}{2} (H \tan \theta)^{-2} \right], \label{eq:ne} \\ 
T_{\rm e} &= T_{\rm e,0} \left( \frac{r}{r_g} \right)^{-\gamma} , \label{eq:Te}
\end{align}
for a gravitational radius $r_g = GM_{\rm BH}/c^2$, and we used the standard Boyer-Lindquist coordinate system $(t,r,\theta,\phi)$ in Kerr spacetime of a rotating BH, parameterized with a spin (angular momentum) parameter $a$ \citep{Kerr1963}. We also defined a dimensionless spin parameter $a_* = a/M$. Parameter $H$ (dimensionless as well) controls the geometric thickness of the disk \citep{2018ApJ...863..148P, Vincent2022}. We took $H=0.5$ as a fiducial geometrically thick model and additionally tested $H=0.1$ and $H=0.3$. There are some observational handles on $\gamma$ in RIAF sources based on radiative spectra \citep[e.g.,][]{Broderick2011} and VLBI brightness temperatures \citep{Wielgus2024} of Sgr~A*, and we therefore adopted a value of $\gamma=0.84$ for our fiducial model, which is consistent with observational constraints. Our choices for all fiducial model parameters are given in Table~\ref{table:param}.

To model the four-velocity of the accretion flow, we followed \citet{Pu2016}, \citet{Tiede2020}, and many others and interpolated between Keplerian orbital motion and geodesic free fall, hence
\begin{align}
    u^r &= u^r_{\rm K} + \kappa_{\rm ff}(u^r_{\rm ff} - u^r_{\rm K}),\\
    \Omega &= \Omega_{\rm K} + (1-\kappa_{\rm K})(\Omega_{\rm ff} - \Omega_{\rm K}),
\end{align}
where ($u^r_{\rm K}, \Omega_{\rm K}$) and $(u^r_{\rm ff}, \Omega_{\rm ff})$ represent the radial four-velocity component $u^r$ and angular velocity $\Omega = u^\phi/u^t$ for Keplerian and for free-fall motion with zero angular momentum, respectively. In the Keplerian case, fluid motion inside the innermost stable circular orbit (ISCO) is modeled as a radial inspiral, following from the conservation of the ISCO values of energy $E = -u_t$ and angular momentum $L = u_\phi$. We set $u^\theta = 0$ and obtained $u^t$ from the four-velocity normalization. We assumed $u^r < 0$, that is, we did not allow for outflows (hence, no emission from the jets). Under the assumed parameterization, Keplerian motion corresponds to $(\kappa_{\rm ff},\kappa_{\rm K}) = (0,1)$, and free fall corresponds to $(\kappa_{\rm ff},\kappa_{\rm K}) = (1,0)$. Fractional values of these parameters result in a mixed nongeodesic prescription that is more appropriate for a model of a realistic astrophysical flow in \m87, where a sub-Keplerian azimuthal velocity profile and advection are expected \citep[e.g.,][]{2011ApJ...729...86T}. We considered $(\kappa_{\rm ff},\kappa_{\rm K}) = (0,1)$,  $(1,0)$ and $(0.5,0.5)$, with the latter used as the fiducial RIAF model.
\begin{figure}
    \centering
    \includegraphics[width=\columnwidth]{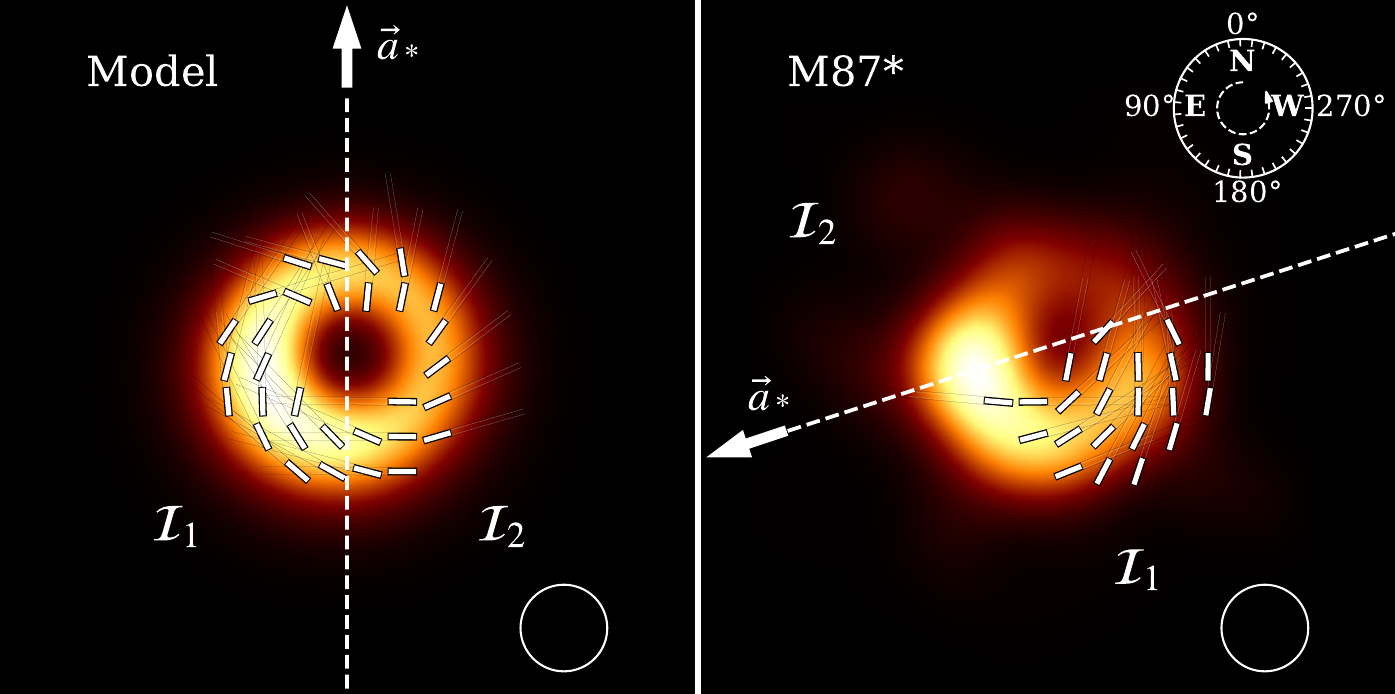}
    \caption{Left: Example of an RIAF model with $a_* > 0$ ray-traced at an inclination of $163^\circ$, blurred with a $15\, \muas$ circular Gaussian, indicated with a circle in the bottom right corner, before the 108$^\circ$ counterclockwise rotation to match it with the observed appearance of the \m87. The ticks indicate the direction of polarization, vector $\vec{a}_*$ shows the observer's screen projection of the BH spin vector pointing away from the observer (into the screen). Right: Image of \m87 reconstructed from the EHT 2017 observations \citep{2019ApJ...875L...4E}. The axis shows the observed position angle of the jet, with the forward jet appearing toward the west at about 288$^\circ$ \citep{Walker2018}. The arrow shows the expected direction of the BH spin projection, pointing away from the observer. The $15\, \muas$ 
 effective resolution of the image is indicated with a circle in the bottom right corner. Polarization ticks are shown in the region where $\mathcal{I}>0.1\mathcal{I}_{\rm max}$ and $\vert\mathcal{P}\rvert=\sqrt{\mathcal{Q}^2 + \mathcal{U}^2} >0.2\lvert\mathcal{P}\rvert_{\rm max}$, as shown in Fig.~1 of \cite{2021ApJ...910L..13E}. }
    \label{fig:ratio}
\end{figure}
\begin{figure*}[h!]
    \centering
    \includegraphics[width=0.78\textwidth]{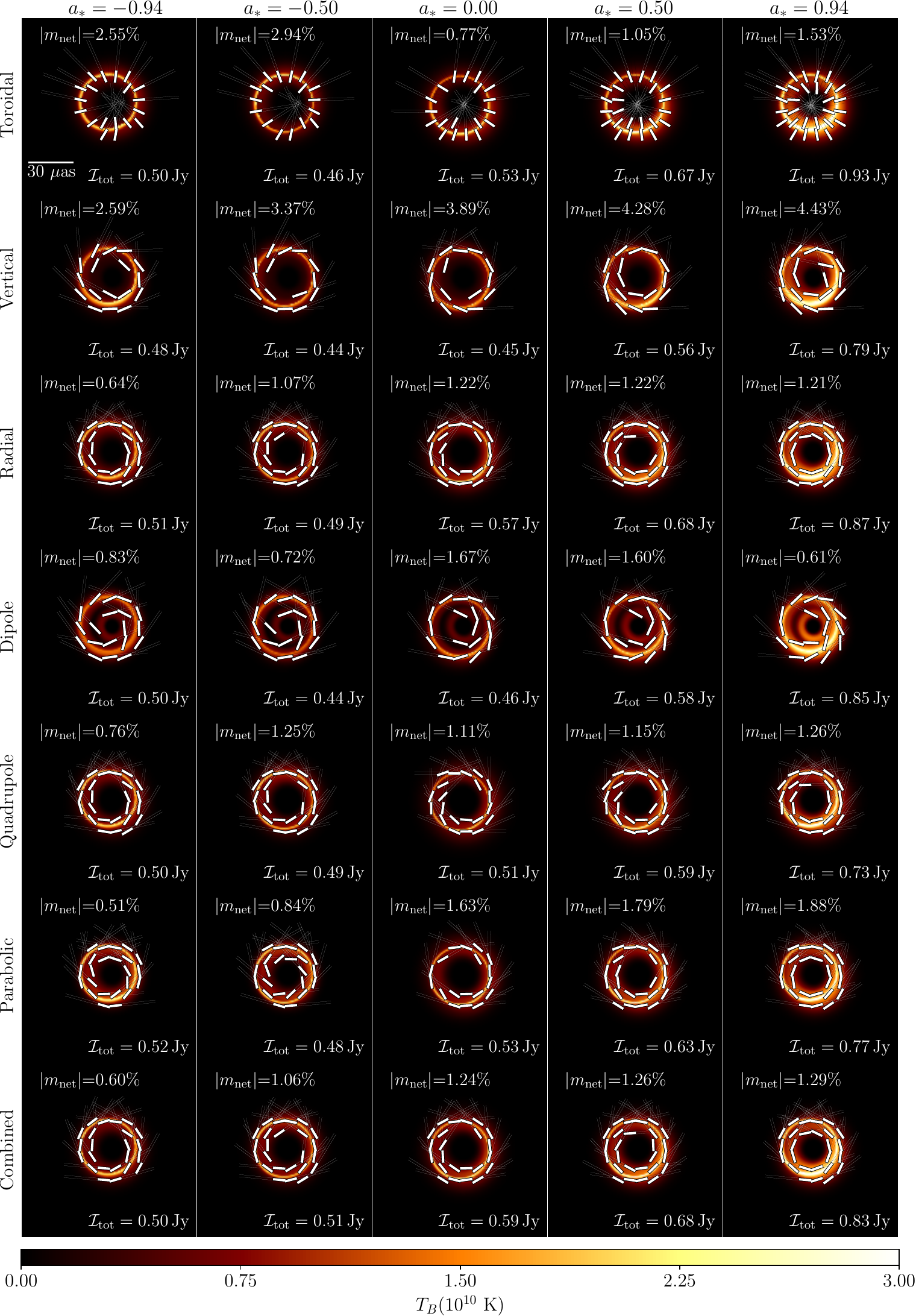}
    \caption{Ray-traced images of the fiducial RIAF model (Table~\ref{table:param}) with various magnetic field geometries and different BH spin $a_*$ values, shown in units of brightness temperature. The conventions for the inclination angle $\theta_i$ and the position angle $\theta_{\rm PA}$ are described in Sect.~\ref{subsec:emission}. The total flux density $\mathcal{I}_{\rm tot}$ and the net fractional polarization $|m_{\rm net}|$ are indicated. The convention for showing EVPA ticks is the same as in Fig.~\ref{fig:ratio}. }
    \label{fig:raytrace_inc_163}
\end{figure*}

\begin{figure*}
    \centering
    \includegraphics[width=0.78\textwidth]{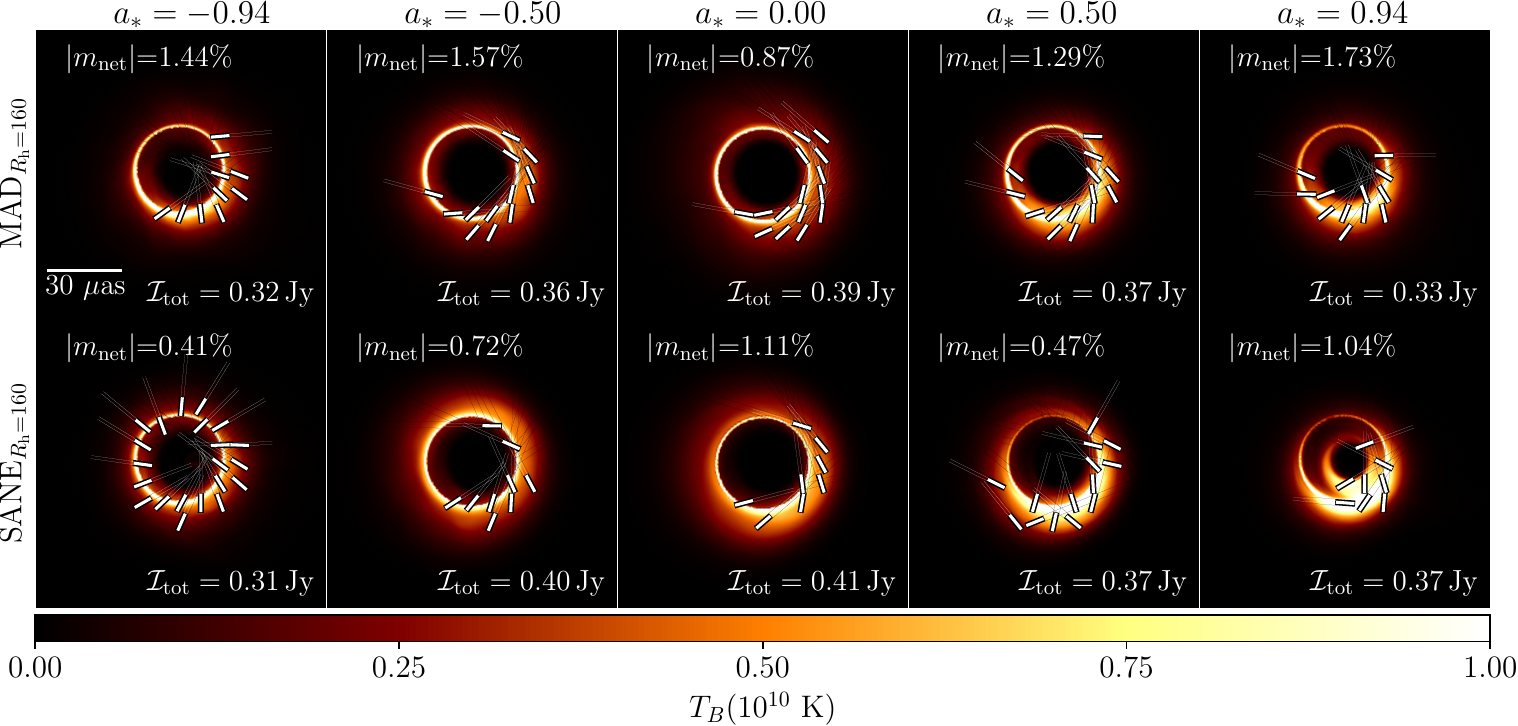}
    \caption{Time-averaged ray traced GRMHD images from the EHT image library \citep{2019ApJ...875L...5E,2021ApJ...910L..13E,Wong2022}, representing different BH spin values $a_*$ and MAD or SANE states of magnetization. The conventions for the inclination angle $\theta_i$ and the position angle $\theta_{\rm PA}$ are described in Sect.~\ref{subsec:emission}. The convention for showing EVPA ticks is the same as in Fig.~\ref{fig:ratio}.}
    \label{fig:grmhd}
\end{figure*}
\subsection{Magnetic field geometry}\label{sec:magnetic_fields}

We considered several geometries of the magnetic field, embedded in the accretion flow described in Sect.~\ref{sec:flow_description}. These are typically defined through vector potential $\vec{A}$, thus enforcing the vanishing divergence of the magnetic field $\nabla_i B^i = 0$. In our convention, Latin indices correspond to three spatial dimensions, while Greek ones span over all four spacetime dimensions.
We assumed axisymmetric configuration, with the only nonzero component $A_\phi$, that is, $A_r=A_\theta= 0$. The frame–dragging effect twists magnetic field lines and inevitably induces an electric field. As shown by \cite{2004MNRAS.350..427K,2022MNRAS.512.2798K}, this induced electric field cannot be completely screened in the vicinity of a rotating BH. In particular, it is unscreened within the ergosphere. This is a general feature of this class of semi-analytic models, and we only focused on the magnetic field effect on the observables here. We considered the following seven geometries, illustrated in Fig.~\ref{fig:bfield_schematic}:

\begin{itemize}
    \item Radial \citep[split monopole, e.g.,][]{1977MNRAS.179..433B}
         \begin{equation}
         A_{\phi, R} =  1 - |\cos{\theta}|,
        \end{equation}
    \item vertical  \citep[e.g.,][]{2022A&A...668A.185V}
         \begin{equation}
         A_{\phi, V} =  r \sin{\theta},
        \end{equation}
    \item dipole \citep{2009MNRAS.394L.126M}
        \begin{equation}
        A_{\phi, D} =  \frac{1}{2}((r + r_0)^\nu\mathcal{F}_{-} + 2 \mathcal{F}_+(1 - \ln(\mathcal{F}_+))),
        \end{equation}
     where $\mathcal{F}_-= 1-\cos^\mu{\theta}$, $\mathcal{F}_+ = 1+\cos^\mu{\theta}$, $\nu = 3/4$, $\mu = 4$, and $r_0=4r_g$,
    \item quadrupole \citep{2009MNRAS.394L.126M}
        \begin{equation}
        A_{\phi, Q} = A_{\phi, D} \cos{\theta},
        \end{equation}
    \item parabolic \citep[e.g.,][]{SashaT2010,2023EPJC...83..323K}
        \begin{equation}
        A_{\phi, P} = r^k (1 - |\cos{\theta}|) = r^k A_{\phi, R},
        \end{equation}
        where we set $k=0.75$ and for $k=0$, this expression reduces to the radial magnetic field (split monopole),
    \item combined field \citep{2024PhRvD.109f3005K}. This solution combines two fundamental field configurations, namely the vertical homogeneous solution of \citet{Wald:1974:PHYSR4:} and the radial (split-monopole), resulting in a paraboloidal field that is a self-consistent solution to the Maxwell equations in Kerr spacetime,
    
    \begin{equation}
       A_{\phi, C} = g_{\phi\phi} + 2 a g_{t\phi} + p_0\frac{(r^2 + a^2)|\,\cos{\theta}\,|}{r^2 + a^2 \cos^2{\theta}},  \label{eq:combined}
    \end{equation}
    where $p_0$ is the ratio of the radial to vertical magnetic field components.
    \item Finally, for a toroidal configuration, it is straightforward to directly define a divergence-free magnetic field as 
        \begin{equation}
          \ B^r = B^\theta = B^t = 0, \  B^\phi = \frac{1}{\sqrt{-g}} \ , 
        \end{equation}
    where $g$ is the Kerr metric determinant.
\end{itemize}
For the configurations defined with a vector potential, with $A_r = A_\theta = 0$, the laboratory frame (static observer's frame) magnetic field components can be defined simply as
 \begin{equation}
           B^r =  \frac{\partial_\theta A_\phi }{\sqrt{-g}}\ \ , \ \ B^\theta =  -\frac{\partial_r A_\phi }{\sqrt{-g}}\ , \ \ B^\phi = B^t = 0 \ .
\end{equation}
The asymptotic behavior of the magnetic field strength following these prescriptions is $|B| \propto r^{-1}$ for the vertical and toroidal configurations, $|B| \propto r^{k-2}$ for the parabolic field, and $|B| \propto r^{-2}$ for the case of a radial field, dipole and quadrupole. The combined field asymptotes to a constant vertical magnetic field. Furthermore, the dipole field is predominantly vertical in the equatorial plane, and the quadrupole is predominantly radial in the equatorial plane, which is relevant for the signatures of a geometrically thin disk.

 Subsequently, we normalized the magnetic fields by a constant factor, so that the magnetic field strength at the equatorial ring $(r = 3 r_g, \theta = \pi/2)$ was equal to $B_0=5$\,G, which is comparable to the EHT estimates \citep{2021ApJ...910L..13E}. In ideal magnetohydrodynamics the laboratory frame magnetic field $B^\mu$ is related to the fluid-frame magnetic field four-vector $b^\mu$ as \citep{gammie2003}
 \begin{equation}
           b^t = B^iu_i \ \  , \ \ b^i = \frac{B^i + b^tu^i}{u^t} \ ,
\end{equation}
for the fluid four-velocity $u^\mu$, which we have defined in Sect.~\ref{sec:flow_description}. The magnitude of $b^\mu$ and its angle with respect to the momentum of the emitted photon $k^\mu$ were then used to compute the intrinsic synchrotron emission, along with other radiative transfer parameters. While the magnetic fields that we defined are divergence-free by construction, we did not enforce the resulting configuration of $(b^\mu, u^\mu)$ to be a stationary solution of the full induction equation in Kerr metric.
\subsection{Emission and synthetic images}\label{subsec:emission}

In order to perform the radiative transfer and ray-trace the synthetic images of the source, we followed the numerical setup described by \citet{2018MNRAS.475...43M}. At radio frequencies, the emission from RIAF systems is dominated by the synchrotron radiation produced by hot electrons gyrating in the magnetic field \citep{2014ARA&A..52..529Y}. We assumed a relativistic thermal (Maxwell-J\"{u}ttner) energy distribution of electrons and followed the fitting formulas of \citet{2016ApJ...822...34P} to compute the necessary coefficients of the fully polarized radiative transfer, including emission, absorption, and Faraday effects.

 For the mass and distance of \m87, we assumed $M_{\rm BH} = 6.4\times10^9 \rm{M}_\odot$ and $D_{s} = 16.9\,\rm{Mpc}$, respectively \citep{2011ApJ...729..119G, 2019ApJ...875L...6E,2025A&A...693A.265E}. We set the observing frequency to $\nu_{\rm obs} = 230\,$GHz, which is within the observing frequency band of the EHT \citep{2019ApJ...875L...2E,2019ApJ...875L...3E}. The inclination (viewing angle) with respect to the jet axis $\theta_i$ was estimated observationally \citep{Mertens2016,Walker2018}, with the ambiguity between $\theta_i$ and $180^\circ - \theta_i$ following from the overall symmetry of the system. The symmetry is broken by the orientation of the rotation, however. We adopted $\theta_i = 17^\circ$ or $\theta_i = 163^\circ = 180^\circ - 17^\circ$, assuming that the jet axis, angular momentum vector of the BH, and that of the accretion flow are all aligned along the same axis. The sign of the BH spin $a_*$ informs us whether the accretion flow and BH angular momenta are parallel (prograde rotation, $a_* > 0$) or antiparallel (retrograde accretion, $a_* < 0$). For our RIAF model, which lacks a jet, we identified $\theta_i$ with the angle calculated with respect to the angular momentum vector of the accretion disk rather than that calculated with respect to the angular momentum of the BH (which is either parallel or antiparallel). An additional prior comes from the EHT results \citet{2019ApJ...875L...1E}, which constrain the brighter side of the observed ring to be located in the bottom side of the image. \citet{2019ApJ...875L...5E} argued that reproducing this orientation within the GRMHD library of images requires that the angular momentum vector of the BH is pointed away from the observer's screen (see also comments in Sect.~\ref{subsec:ring_asymmetry}). Thus, we followed the convention of the EHT and took $\theta_i = 163^\circ$ for the BH spins $a_* \ge0$ and $\theta_i = 17^\circ$ for $a_* < 0$, which guarantees that the BH spin vector points away from the observer in all cases. Keeping $\theta_i = 163^\circ$ for all BH spin values would result in a bright side of the ring flipped to the upper part of the image for $a_* < 0$ GRMHD models, which is inconsistent with the EHT results.

 We ray-traced images in the field of view (FOV) of $200\,\mu$as, corresponding to about $40 \,r_g/D_{s}$. The emission in the wider FOV in our idealized RIAF disk model is negligible, similarly as in the images reconstructed by the EHT \citep{2019ApJ...875L...4E, 2024A&A...681A..79E}. For each considered magnetic field model, accretion flow parameters $\kappa_{\rm ff}$, $\kappa_{\rm K}$, and disk thickness parameter $H$, we tuned the number density scale $n_{\rm e,0}$ and temperature scale $T_{\rm e,0}$ to match the typically observed compact scale emission in \m87 of $\mathcal{I}_{\rm tot} =$ 0.5\,Jy at 230\,GHz \citep{2019ApJ...875L...4E,Wielgus2020, 2024A&A...681A..79E} within 10\% for the model with dimensionless spin $a_*=-0.94$. We kept the same normalization for other spin values, which led to a flux density deviation from 0.5\,Jy of up to a factor of 2. In all cases, our $n_{\rm e,0}$ was within an order of magnitude from $10^4$\,cm$^{-3}$, and $T_{\rm e,0}$ was within an order of magnitude from $10^{11}$\,K. This is reasonably consistent with the EHT estimates \citep{2021ApJ...910L..13E}.

 \subsection{GRMHD images}
 \label{subsec:GRMHD}
 
For comparison purposes, we also considered time-averaged GRMHD images from the EHT 230\,GHz images library \citep{2021ApJ...910L..13E,Wong2022, Dhruv2025}. In these simulations, plasma and magnetic field co-evolve following the conservative MHD equations in a curved Kerr spacetime, reaching a quasi-stationary turbulent state of accretion. We considered ten simulations, computed for five different values of BH spin and for two different states of magnetization, a magnetically arrested disk (MAD), characterized by strong, coherent magnetic fields \citep{Narayan2003}, and a weak, turbulent magnetic field solution, ``standard and normal evolution'' \citep[SANE; ][]{2012MNRAS.426.3241N}. Each of these two states predicts distinct dynamical outcomes and observable phenomena \citep{2011MNRAS.418L..79T,2012MNRAS.426.3241N}. The models differ primarily in the strength and configuration of the magnetic fields threading the accretion flow, leading to distinct predictions for accretion rates, jet power, and observable emission characteristics. In the MAD regime, the magnetic flux threading the BH accumulates until it reaches a saturation point, which results in a magnetically dominated inner accretion flow. The accretion flow becomes highly nonaxisymmetric, with low-density magnetic flux bundles alternating with high-density streams of gas  \citep{2008ApJ...677..317I}, and the magnetic field lines exhibit a characteristic split-monopole configuration in the inner regions \citep{2011MNRAS.418L..79T}. In contrast, the SANE model represents a weakly magnetized accretion flow in which the accretion flow maintains a more axisymmetric structure than in MAD \citep{2012MNRAS.426.3241N}. In this case, the magnetic field lines exhibit a more tangled and turbulent configuration \citep{2014MNRAS.439..503S}. The location of the compact scale emission observed by the EHT in \m87 was discussed in \citet{2019ApJ...875L...5E}. All MAD models considered in the EHT library were found to  dominated by the near-equatorial emission from the accretion flow. Emission from the jet base only manifests itself in some of the SANE models under assumptions resulting in a relatively low temperature of the electrons in the accretion disk.

For all models we considered, the electron energy distribution was relativistic thermal, which is a common assumption for modeling emission at frequencies $\nu_{\rm obs}$ relevant for this work \citep{2019ApJ...875L...5E}. While the electron temperature $T_{\rm e}$ in our RIAF models is governed by a radial power law given by Eq.~\ref{eq:Te}, however, the GRMHD models prescribe $T_{\rm e}$ as a function of the ion temperature $T_{\rm i}$ and the plasma $\beta_{\rm p}$ parameter, a local ratio of gas pressure to magnetic pressure. These two quantities are constrained within a GRMHD simulation, and the local ion-to-electron temperature ratio is then evaluated as a postprocessing step with a parameter $R_{\rm h}$, following \citet{Moscibrodzka2016}
\begin{equation}
\frac{T_{\rm i}}{T_{\rm e}} =  \frac{ R_{\rm h} \beta_{\rm p}^2 + 1}{1 + \beta_{\rm p}^2} \ .
\end{equation}
The results of \citet{2019ApJ...875L...5E,2021ApJ...910L..13E,M87Paper9}, and \citet{Joshi2024}, where a range of values $1 \leq R_{\rm h} \leq 160$ were investigated, indicate a preference for a high value of $R_{\rm h}$. In particular, most SANE models are inconsistent with a rather conservative lower limit on the observed jet power with only high negative BH spin or high positive BH spin, and $R_{\rm h} \geq 80$ surviving this constraint \citep{2019ApJ...875L...5E}. Hence, for comparisons with our RIAF models, we adopted a fixed value of $R_{\rm h} = 160$, resulting in relatively cold electrons in the accretion disk region, and as a consequence, in stronger Faraday effects \citep{Quataert2000} and associated depolarization \citet{2021ApJ...910L..13E,SgrA_P8_2024}. Another consequence of larger $R_{\rm h}$ is that in order to reproduce the observed total intensity with colder electrons, a higher mass-accretion rate is required, which further increases the Faraday depth and generally enhances the jet power.

\begin{table} 
    \centering
    \caption{EHT constraints on \m87 \citep{2019ApJ...875L...1E,2021ApJ...910L..12E,2021ApJ...910L..13E,2024A&A...681A..79E}.}
    \begin{tabular}{c c c}
        \hline
        Parameter & Lower Bound & Upper Bound \\
        \hline
        $\mathrm{\alpha}$ & $10.3$ & $11.7$\\
        $\mathcal{I}_r$ & $0.55$ & $0.75$\\
        $\lvert m \lvert_{\rm net} $ & $1\%$ & $3.7\%$ \\
        $ \langle\lvert m \rvert  \rangle$ & $5.7\%$ & $10.7\%$ \\
        $\lvert \beta_2 \rvert$ & $0.04$ & $0.07$ \\
        $\angle \beta_2$ & $-163^\circ$ & $-127^\circ$ \\
        $\mathrm{EVPA}$ & $0^\circ$ & $-80^\circ$ \\
        \hline
    \end{tabular}
    \label{table:constraints}

\end{table}

\section{Method} \label{sec:3}

To compare our synthetic images of the RIAF model and the EHT observations of \m87  \citep{2019ApJ...875L...1E, 2021ApJ...910L..12E}, we defined quantifiable image metrics (not to be confused with the metric tensor) characterizing the source appearance in observations and in model images. We discuss the angular diameter of the ring and its brightness asymmetry as total intensity image metrics. Furthermore, we considered polarimetric consistency metrics following the method of \citet{2021ApJ...910L..13E}. The summary of the EHT constraints for the image metrics we considered is given in Table~\ref{table:constraints}.
\subsection{Angular diameter of the ring} \label{sec:diameter}
We extracted the size of the ring from the images using the \texttt{REx} module in the \texttt{eht-imaging}\footnote{\url{https://github.com/achael/eht-imaging}} library \citep{Chael2016}, following the procedures of \citet{2019ApJ...875L...4E}. The \texttt{REx} module allowed us to calculate the radial profiles of the intensity distribution, which can be used to determine the center and diameter of the ring. The diameter $\mathcal{D}$ was estimated by the radial distance of the pixel with the peak intensity in the images. We blurred the synthetic RIAF images with a 15\,$\mu$as circular Gaussian before extracting the profiles, in order to mimic the effective resolution of the EHT. Since the observable angular diameter scales with the uncertain mass over distance estimate $\theta_g = r_g/D_{s}$, equal to $3.74$\,$\mu$as for our fiducial parameters, we report the following normalized parameter \citep{2019ApJ...875L...6E}:
\begin{equation}
    \alpha = \frac{\mathcal{D}}{\theta_g} = \frac{\mathcal{D}}{3.74 \,\mu{\rm as}} \ .
\end{equation}
The EHTC reported a measurement of parameter $\alpha$ for \m87 to be $\alpha = 11.0 \pm 0.7$, corresponding to a diameter of $42\pm 3~\mu$as under the assumed $r_g/D_{s}$ \citep{2019ApJ...875L...1E,2024A&A...681A..79E}. This is remarkably close to the theoretical prediction for the geometric BH shadow, in the sense of a trace of the photon shell \citep[e.g.,][]{Bardeen1972} on the distant observer's screen (the ``apparent boundary'' of \citealt{Bardeen1973} or the ``critical curve'' of \citealt{Gralla2019}). This prediction, which is dictated exclusively by GR through the null geodesic structure of spacetime, gives $\alpha = 6 \sqrt{3} \approx 10.4$ for a Schwarzschild BH.

\subsection{Brightness asymmetry} \label{sec:3.1}
To quantify the brightness asymmetry in the total intensity image, we considered the ratio of the observed flux density between two halves of the image, separated by the line of the projected spin axis as indicated by Fig.~\ref{fig:ratio},
\begin{equation}
    \mathcal{I}_r = \frac{\mathcal{I}_2}{\mathcal{I}_1} \ .
\end{equation}
We used \texttt{REx} in order to define the center of the ring. We used the position angle of the forward jet, constrained by the lower-frequency radio observations to be $\theta_{\mathrm{PA}} = 288^\circ$ \citep{Walker2018}, as a proxy for the BH spin axis (see Fig.~\ref{fig:ratio} and comments in Sect.~\ref{subsec:emission}). We measured $\mathcal{I}_r$ in the 2017 EHT images of \m87 \citep{2019ApJ...875L...4E} to be 0.7$\pm$0.1 and in the 2018 EHT images of \m87 \citep{2024A&A...681A..79E} to be 0.6$\pm$0.1, and we therefore took 0.65$\pm$0.10 as the measurement of $\mathcal{I}_r$ to be compared to the models. 

\subsection{Polarimetric observables}\label{sec:3.2}

We followed \citet{2021ApJ...910L..12E,2021ApJ...910L..13E} to define the metrics quantifying the polarization properties in the resolved and unresolved images. We denote the flux densities integrated over the field of view solid angle $A$ as $\mathcal{I}_{\rm tot}$ for the total intensity,
\begin{equation}
\mathcal{I}_{\rm tot}=   \int_A \widetilde{\mathcal{I}} d A  \\,
\end{equation}
and $\mathcal{P}_{\rm tot}$ for the linear polarization,
\begin{equation}
\mathcal{P}_{\rm tot}=   \int_A \widetilde{\mathcal{P}} d A =   \int_A \left( \widetilde{\mathcal{Q}} + i \widetilde{\mathcal{U}} \right) d A \ .
\end{equation}
$\mathcal{I}_{\rm tot}$ and $\mathcal{P}_{\rm tot}$ are measured in jansky (Jy), while quantities with a tilde refer to the flux density per solid angle (brightness). $\mathcal{I}$, $\mathcal{Q}$, $\mathcal{U}$, $\mathcal{V}$ are the four Stokes parameters, describing the observed state of polarized emission. $\mathcal{P}_{\rm tot}$ is a complex number defining the strength and orientation of the polarized emission. The net polarization fraction, corresponding to a coherent average of $\widetilde{\mathcal{P}}$ over the field of view and thus measurable from unresolved observations, is defined as
\begin{equation}
\lvert m \rvert _{\rm net} =   \frac{ \left| \mathcal{P}_{\rm tot}  \right| }{ \mathcal{I}_{\rm tot}  } = \frac{1}{ \mathcal{I}_{\rm tot}  }  \left| \int_A \widetilde{\mathcal{P}} d A \right| \ .
\end{equation}
For resolved EHT images, we also defined the incoherently averaged linear polarization fraction 
\begin{eqnarray}
    \langle\lvert m \rvert  \rangle = \frac{1}{ \mathcal{I}_{\rm tot}  }  \int_A  \left| \widetilde{\mathcal{P}} \right| d A  \ .
\end{eqnarray}
This metric depends on the effective image resolution, and following \citet{2021ApJ...910L..12E}, we computed it on images blurred with a 15\,$\mu$as circular Gaussian to mimic the effective EHT resolution. Furthermore, we considered the second mode of the azimuthal (in the $\phi$ angle direction) Fourier decomposition of the polarized image, denoted as $\beta_2$ \citep{2020ApJ...894..156P},
\begin{equation}
    \beta_2 = \frac{1}{  \mathcal{I}_{\rm tot}  } \int_A \widetilde{\mathcal{P}} \,e^{-2 i \phi}\, dA \ .
\end{equation}
The $\beta_2$ metric exploits the approximate polar symmetry of the BH image observed at low inclination $\theta_i$. The argument of this complex number characterizes the angle between the EVPA pattern and the BH image ring, while its absolute value informs us about how rotationally symmetric the EVPA pattern is.

\begin{figure}
    \centering
    \includegraphics[width=1.0\columnwidth]{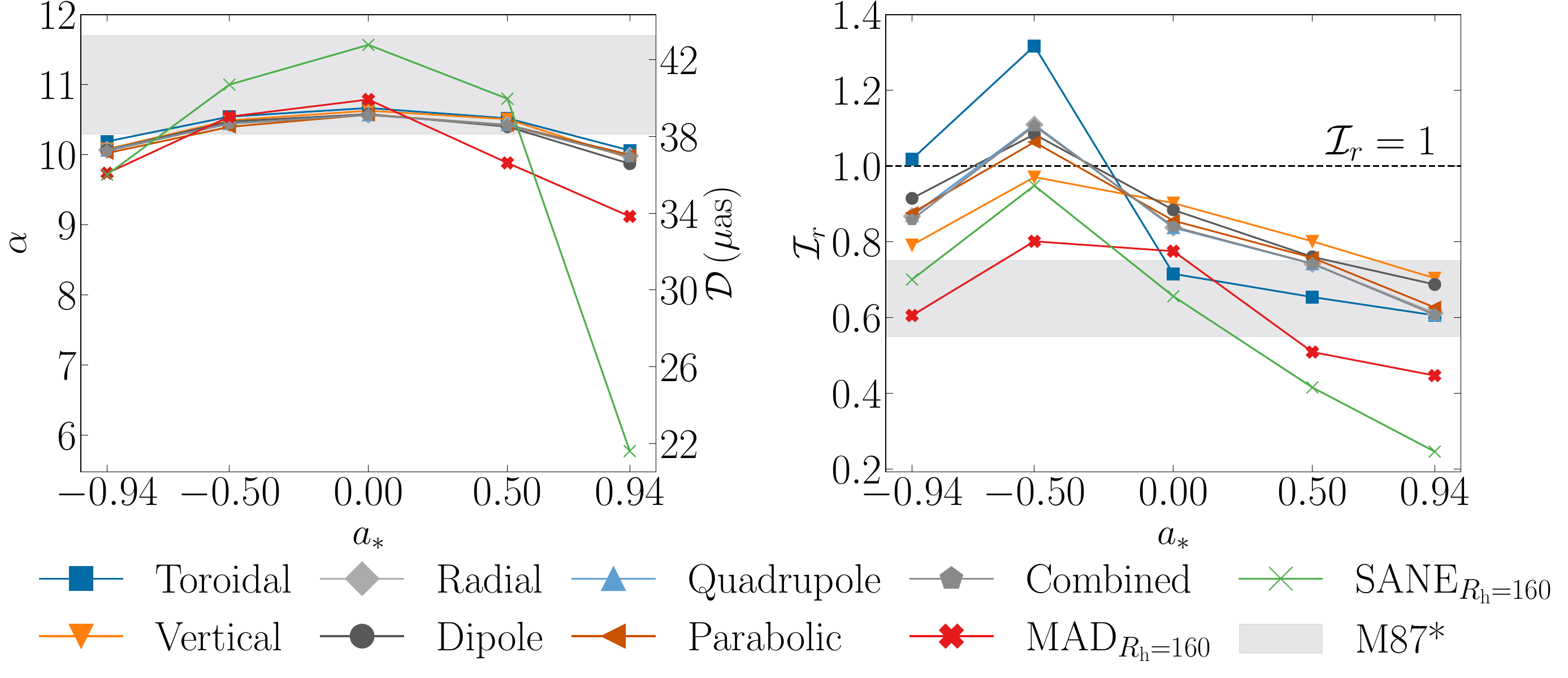}
    \caption{Left: Diameter of the ring $\mathcal{D}$ and normalized diameter $\alpha$ for the fiducial RIAF model with various magnetic field configurations and for GRMHD images as a function of BH spin $a_*$. Right: Same for the brightness ratio parameter $\mathcal{I}_r$. The gray bands indicate constraints derived from the EHT observations of \m87.}
    \label{fig:intensity_ratio_subkep_bfield}
\end{figure}

\begin{figure*}
    \centering
    \includegraphics[width=\textwidth]{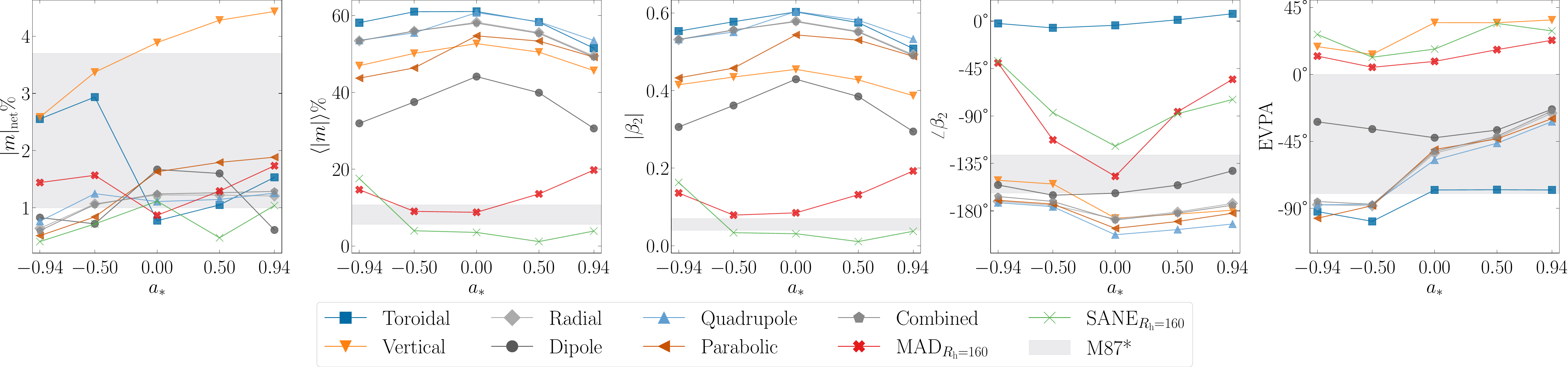}
    \caption{Variation in the net fractional polarization $\lvert m \rvert_{\rm net}$, resolved fractional polarization $\langle\lvert m \rvert\rangle$, amplitude and phase of the $\beta_2$ parameter, and the net $\mathrm{EVPA}$ for the fiducial RIAF model with various magnetic field configurations and for GRMHD images as a function of BH spin $a_*$. The gray bands indicate constraints derived from the EHT observations of \m87.}
    \label{fig:pol_subkep_all}
\end{figure*}
\section{Results}\label{sec:4}

We present a grid of RIAF model images in Fig.~\ref{fig:raytrace_inc_163}. They were calculated for the fiducial parameters given in Table~\ref{table:param} and incorporate the various magnetic field geometries discussed in Sect.~\ref{sec:magnetic_fields} and different values of the BH spin $a_*$. All images exhibit a general ring morphology, with a prominent narrow photon ring \citep{Johnson2020,Paugnat2022}, formed by photons executing at least half loop around the BH on their way to the observer's screen. The reconstructed EHT images, achieving an effective resolution of $\sim 15 \, \mu$as, generally blend this feature with the diffused direct emission (see Fig.~\ref{fig:ratio}). The morphology of RIAF images is very similar to that found in time-averaged GRMHD images shown in Fig.~\ref{fig:grmhd}. One outstanding feature is the second smaller ring that can be seen in SANE $R_{\rm h} = 160$, $a_* = 0.94$, which is the model for which the jet base emission dominates the disk emission \citep{2019ApJ...875L...5E}. A similar feature appears in the RIAF models for a dipole magnetic field configuration, where it is connected to the enhanced emission in the off-equatorial high-density region of magnetic field lines at intermediate values of $\theta$ (see panel (c) of Fig.~\ref{fig:bfield_schematic}). Another apparent feature in the GRMHD images is the depolarization of the northeast part of the ring seen in many of the images. This effect has been attributed to the Faraday rotation scrambling the polarization along light rays that spend a longer time in the accreting plasma \citep{2021ApJ...910L..13E}. The depolarization does not affect our fiducial RIAF images, which generally remain Faraday-thin. This is discussed further in Sect.~\ref{sec:5}.

 \citet{2021ApJ...910L..13E} and \citet{Narayan2021} discussed the relation between the magnetic field geometry and the observable EVPA pattern of the accretion system modeled as a thin emission ring. Following a simple argument of perpendicularity between polarization vector $\vec{P}$, magnetic field $\vec{B}$, and wave vector $\vec{k}$ for synchrotron emission, expressed with a $\vec{P} \propto \vec{k} \times \vec{B}$ relation, they concluded that toroidal magnetic fields produce a radial EVPA pattern and poloidal magnetic fields generally produce azimuthal EVPA patterns. With the images shown in Fig.~\ref{fig:raytrace_inc_163}, we confirm that these conclusions hold well for a more realistic spatially extended global emission models and geometries, as long as the Faraday depth is low.

\subsection{Angular diameter of the ring}
\label{subsec:ring_size}

\begin{figure*}
    \centering
    \includegraphics[width=\textwidth]{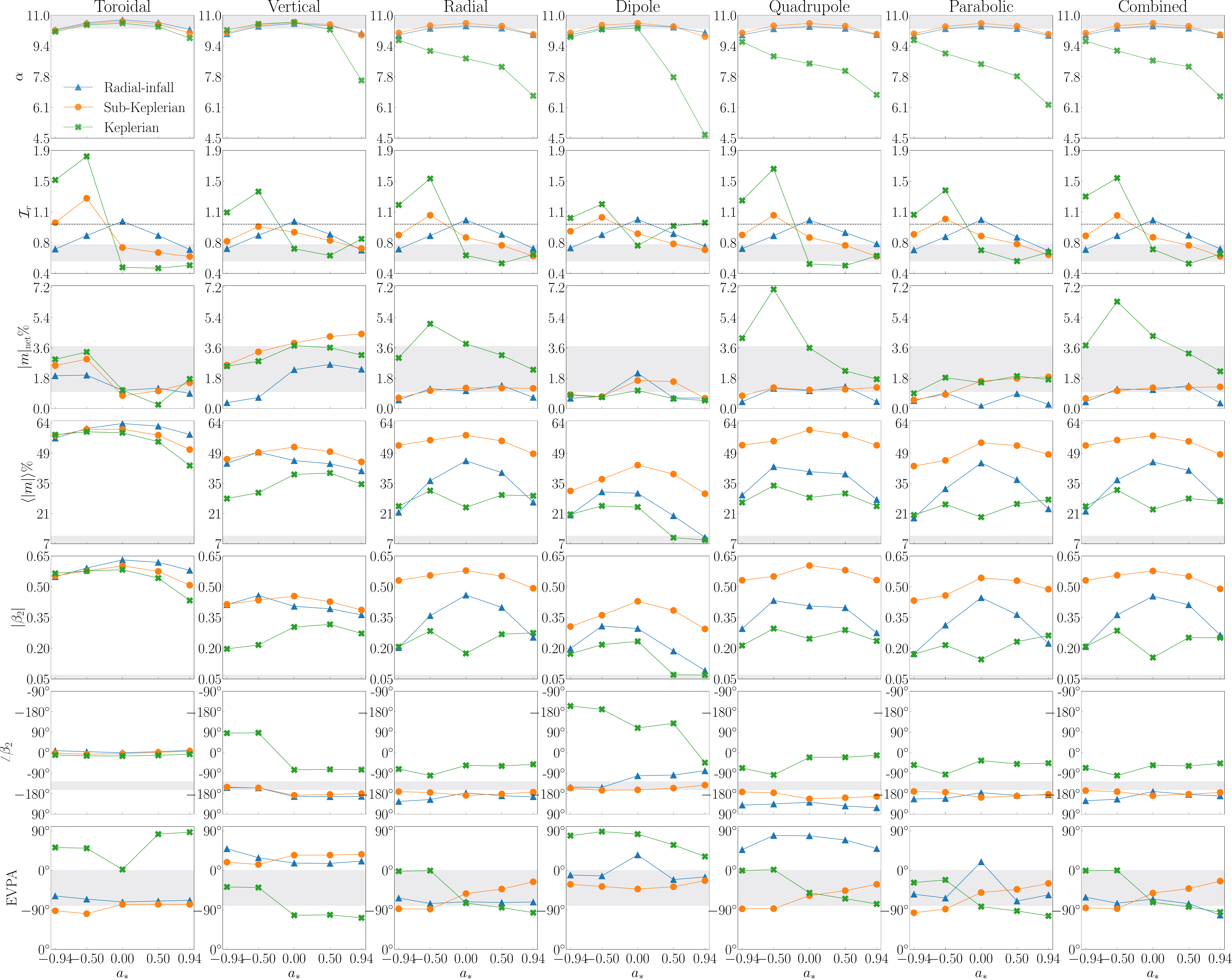}
    \caption{Dependence of the image metrics discussed in Sect.~\ref{sec:3} on the magnetic field geometry and BH spin $a_*$ for the fiducial RIAF model. Three models of plasma four-velocity pattern are considered (see Sect.~\ref{sec:flow_description}). The gray bands indicate the constraints derived from the EHT observations of \m87.
    }
    \label{fig:panel_quant_kappa}
\end{figure*}

The ring size $\alpha$ parameter estimates for the fiducial velocity model are given in the left panel of Fig.~\ref{fig:intensity_ratio_subkep_bfield}. There is a small bias, with most models producing a slightly lower $\alpha$ than the observational constraints of the EHT. This can easily be explained through errors on the assumed a priori $\theta_g = r_g/D_{s}$, however, which are about 10\%. The ring diameter depends weakly on the BH spin seen in the RIAF and in GRMHD simulations, with low spin models forming slightly larger rings. The effect is related to the BH event horizon and BH shadow, in the sense of a trace of the photon shell \citep[e.g.,][]{Bardeen1972} on the distant observer's screen (``apparent boundary'' of \citealt{Bardeen1973} or the ``critical curve'' of \citealt{Gralla2019}) that decreases with the increase in $|a_*|$. 

It is notable that the sign of $a_*$ only plays a secondary role for the image size. The magnetic field configuration, on the other hand, does not significantly affect the measured diameter for the fiducial RIAF model. The high-spin SANE GRMHD model results in a significantly lower measurement of the diameter as a consequence of the jet base emission. This demonstrates that models dominated by jet emission are discouraged under $r_g/D_{s}$ priors and given the EHT observations \citep[][and Sect.~\ref{sec:model} of this paper]{2019ApJ...875L...5E,2021ApJ...910L..13E}. Furthermore, in the first row of Fig.~\ref{fig:panel_quant_kappa}, we inspect the effect of the accretion flow velocity model on the normalized diameter $\alpha$. While the fiducial sub-Keplerian and free-fall four-velocity models are close to the EHT constraints as well as to the GR predictions of the geometric BH shadow, Keplerian models with a high spin produce significantly smaller observable rings, with a size decreasing with $a_*$. In these models, the accretion flow velocity remains purely azimuthal above the ISCO, while below the ISCO, the growing radial inflow velocity component in the plunging region results in beaming radiation toward the BH. This reduces the observed brightness and adds to the effects of gravitational redshift. Thus, most of the observed emission in the images of Keplerian models corresponds to the region close to ISCO. The ISCO radius is very sensitive to (signed) BH spin, changing from $9M$ for $a_* = -1$ to $1M$ for $a_* = 1$ \citep{Bardeen1972}. At the same time, the radius of the image of ISCO changes from about 10 $\theta_g$ to about 2 $\theta_g$ \citep[see, e.g., Fig. 4 of][for the face-on case calculations]{Wielgus2021}. This means that for high positive BH spins, the primary image of ISCO manifests itself as a bright compact emission region contained within the geometric shadow (which itself is weakly dependent on spin). This biases the ring size measurements toward lower values. A similar behavior of Keplerian models has been reported by \citet{Vincent2022}. As a consequence, high positive spin models can only be consistent with the $\theta_g$ prior for a non-Keplerian four-velocity, with significant inflow already above the ISCO. This effect underscores the importance of the astrophysical setup and that the measured ring diameter does not always approximately track the size of the geometric shadow \citep[see also][]{Gralla2019,Wielgus2021,Urso2025}. Finally, the geometric thickness of the disk does not affect the diameter strongly, as demonstrated in Fig.~\ref{fig:panel_quant_h}. 

\subsection{Brightness asymmetry}
\label{subsec:ring_asymmetry}

\begin{figure*}
    \centering
    \includegraphics[width=\textwidth]{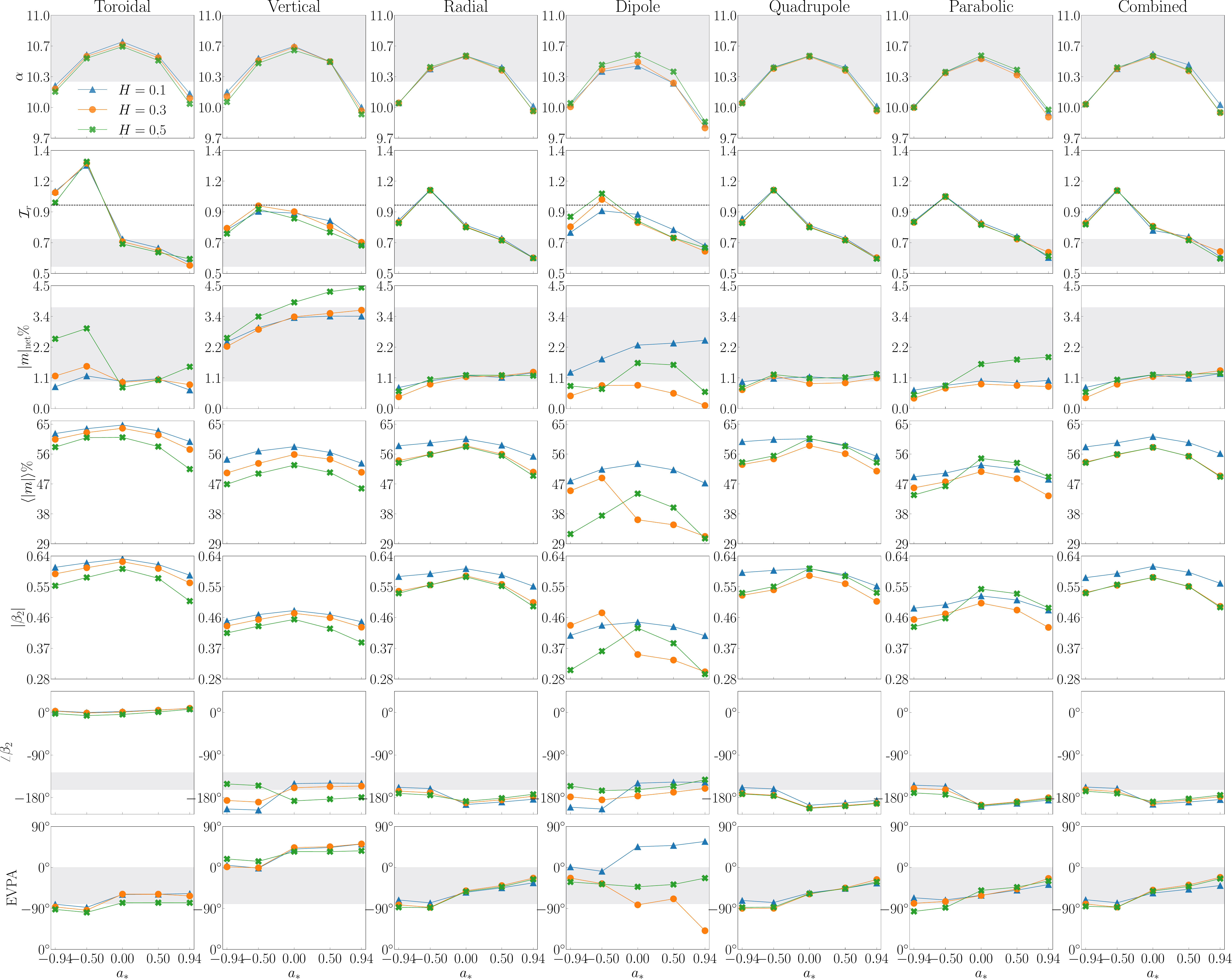}
    \caption{Same as Fig.~\ref{fig:panel_quant_kappa}, but comparing three values of the geometric thickness $H$, as defined in Sect.~\ref{sec:flow_description}.}
    \label{fig:panel_quant_h}
\end{figure*}

The emission forming the 230 GHz \m87 ring is expected to originate from the extremely compact region near the BH event horizon. The EHT observations and their interpretation confirm that most of the observed intensity originates at around 4-5 $r_g$ \citep{2019ApJ...875L...5E}. In this region, the effect of the BH spin on the photon trajectories is significant. Hence, the brightness asymmetry $\mathcal{I}_r$ of the \m87 image essentially originates from the interplay of two effects: special relativistic Doppler boosting associated with the fluid motion, and general relativistic effects related to the BH spin. These effects deform the null geodesics and affect the length of the path that is within the emission region. \citet{2019ApJ...875L...5E} pointed out that for the GRMHD models in the EHT library, the BH spin position controls the direction of the image asymmetry, which is the origin of the convention for the image inclination dependent on the BH spin (see Sect.~\ref{subsec:emission}). Following this convention, if the $\mathcal{I}_r$ estimates presented in the right panel of Fig.~\ref{fig:intensity_ratio_subkep_bfield} are smaller than $\mathcal{I}_r = 1$, then the brighter side of the image is the bottom one ($\mathcal{I}_1$ defined in Fig.~\ref{fig:ratio}). This is consistent with the EHT observations. In particular, for a negative BH spin, $\mathcal{I}_r < 1$ implies that the location of the brighter side is decided by the BH spin and is opposite to what would be observed if the BH did not spin (in which case, the Doppler boost would be the only cause of the brightness asymmetry). For most of our RIAF models, we found $\mathcal{I}_r < 1$ for a negative BH spin. Nonetheless, some Keplerian velocity models with negative BH spin indicate $\mathcal{I}_r > 1$ and might be brought into consistency with the observations by the image position angle rotation (see Fig.~\ref{fig:panel_quant_kappa}). The Keplerian models are also generally most asymmetric, and their brightness asymmetry is driven predominantly by the Doppler effect. Many of these Keplerian models can be rejected based on the excessive amount of asymmetry in comparison to the EHT images. On the other hand, far less asymmetry is observed in RIAF images corresponding to radial infall, for which no Doppler effect is associated with the azimuthal motion (Fig.~\ref{fig:panel_quant_kappa}). As a consequence, the zero angular momentum radial inflow RIAF models are too symmetric and thus inconsistent with the observations for all magnetic field configurations and all BH spins.

With the exception of the negative spin MAD models, GRMHD images appear to be more asymmetric than the RIAF images with the fiducial sub-Keplerian velocity pattern (see Fig.~\ref{fig:intensity_ratio_subkep_bfield}). To obtain the amount of asymmetry similar to the positive spin GRMHD within RIAF framework, we require a more strongly azimuthal velocity pattern and a more azimuthally aligned velocity pattern in conjunction with a toroidal magnetic field  (Fig.~\ref{fig:panel_quant_kappa}). A mixture of toroidal and poloidal magnetic fields is present in the GRMHD models, while the RIAF models we considered are either purely toroidal or purely poloidal.

The dependence of $\mathcal{I}_r$ on the BH spin $a_*$ in Figs.~\ref{fig:intensity_ratio_subkep_bfield}-\ref{fig:panel_quant_kappa} shows that the brightness asymmetry generally increases with growing positive spin because the Doppler and spin effects act in the same direction in this case. For $a_* < 0$, the effects related to BH spin and the Doppler boosting act against one another and in most cases, produce a too symmetric image that is inconsistent with the observational constraints. Hence, the brightness asymmetry constrains the BH spin in a model-dependent way, with a preference toward high positive values of $a_*$ in fiducial RIAF models and moderate positive values for GRMHD. The models with a low negative BH spin are disfavored in both frameworks. Similarly as in case of the ring diameter, the accretion disk thickness $H$ has a very limited effect on the brightness asymmetry (Fig.~\ref{fig:panel_quant_h}). 

\subsection{Polarimetric quantities }
\label{subsec:avg_polar}

For a low inclination angle $\theta_i$, a polarization pattern from an axisymmetric magnetic field averages out to a low value of $|m|_{\rm net}$. Many models, both GRMHD and RIAF, survive this observational constraint (see the first panel of Fig.~\ref{fig:pol_subkep_all}). On the other hand, the resolved polarization fraction $\langle |m| \rangle$ is always too large in our RIAF models, significantly larger than observations and GRMHD models. Many SANE models are even too depolarized, and only some of the MAD models for low or mildly negative BH spin match the observational constraints. The reason for the large $\langle |m| \rangle$ in our RIAF models is likely the low contribution from Faraday effects, which causes significant depolarization in GRMHD models (particularly for a high assumed value of $R_{\rm h}$) and, presumably, in the EHT observations \citep{2021ApJ...910L..13E}. For similar physical reasons, the absolute value of $\beta_2$ is high for the circularly symmetric patterns in RIAF models, but much lower in GRMHD images as well as in the EHT observations. Keplerian models generally produce the least symmetric images (Sect.~\ref{subsec:ring_asymmetry}) and yield the lowest $|\beta_2|$. Nonetheless, the argument of $\beta_2$ appears to be a powerful test of the magnetic field geometry, particularly for a distinction between the dominance of the azimuthal magnetic field component, yielding $\arg \beta_2 \approx 0^\circ$, and various poloidal magnetic field configurations, for which $\arg \beta_2 \approx 180^\circ$. These findings also agree with the results obtained by \cite{2023ApJ...950...38E}. GRMHD models reach values in between, which is expected from a mixed magnetic field configuration (see the fourth panel of Fig.~\ref{fig:pol_subkep_all}). Keplerian RIAF models with poloidal magnetic fields exhibit a somewhat similar behavior (Fig.~\ref{fig:panel_quant_kappa}). Unlike GRMHD, these RIAF models do not obey the fluid four-velocity and magnetic field relation, constrained by the ideal MHD induction equation. The observed $\beta_2$ angle is also in the intermediate range $(-163^\circ,-127^\circ)$, indicating either a mixed magnetic field geometry or an impact from Faraday rotation. For vertical magnetic field configurations, we recovered the trends discussed by \citet{Palumbo2025} and proposed as a test of a BH spin. While the trends are there, however, the values obtained with a global RIAF model differ from the predictions of a thin ring model used by \citet{Palumbo2025}, and a different magnetic field configuration may shift the values further or even reverse the trend direction. Finally, net EVPA, another quantity sensitive to the magnetic field geometry, was only weakly constrained by \citet{2021ApJ...910L..12E}, but broadly corresponds to the direction parallel to the jet ($\theta_{\rm PA} = -72^\circ$). We reproduced the observed EVPA with a number of RIAF models of various magnetic field configurations. The GRMHD models we considered, however,  produce small positive net EVPA and are inconsistent with the observations. In a very simplistic (although commonly used) framework, vertical magnetic fields (parallel to the jet axis) would produce a perpendicular net EVPA at about 18$^\circ$, while azimuthal magnetic fields (perpendicular to the jet axis) would produce a net EVPA around -72$^\circ$. This heuristic works reasonably well for infalling and sub-Keplerian RIAF models with an azimuthal or vertical magnetic field structure (last panel of Fig.~\ref{fig:pol_subkep_all}), but the higher azimuthal velocity and more complicated magnetic field structure affect this simple picture in a nontrivial way, however (last row of Fig.~\ref{fig:panel_quant_kappa}). Furthermore, we ignored the potential effect of Faraday rotation. For  the \m87 rotation measure $\sim\,10^5$ rad m$^{-2}$ \citep{Goddi2021}, the argument of $\beta_2$ can be affected by up to about 20$^\circ$ and EVPA by up to about 10$^\circ$.

\subsection{Effect of the geometric disk thickness} 

\label{sec:4.2}

The RIAF flows are expected to be very hot, and as a consequence, to be geometrically thick. Models often make a simplifying assumption of equatorial emission \citep[e.g.,][]{Gralla2019,Paugnat2022,Urso2025}. We tested the effect of geometric thickness on a number of observables. We summarize the results in Fig.~\ref{fig:panel_quant_h}. In most cases, the geometric thickness does not affect the results significantly. Hence, as long as the disk (and not the off-equatorial jet base) dominates the emission, the equatorial emission assumption provides a reasonable approximation of a more realistic thick disk. One exception is the dipolar magnetic field configuration, for which the geometrically thin disk is mostly threaded by a vertical magnetic field, but as the thickness increases, the emission region extends more into the off-equatorial radial magnetic field region (see Fig.~\ref{fig:bfield_schematic}). The effect is particularly strong for polarimetric observables, which are more sensitive to the magnetic field geometry.

\section{Discussion and conclusions} \label{sec:5}

We used a semi-analytic geometric RIAF disk model of BH accretion to study the effect of the system parameters on the seven observables, polarimetric ones in particular, that are used for evaluating theoretical models of \m87 \citep{2019ApJ...875L...5E, 2021ApJ...910L..13E}. We considered five values of BH spin, seven distinct models of magnetic field geometry, three models of the plasma velocity pattern, and three values of the geometric thickness of the accretion flow. We compared our ray-traced synthetic images with the EHT observations as well as with a suwbset of GRMHD simulations from the EHT \m87 library computed at the observing frequency of 230\,GHz. 

We aimed to verify the simplifying assumptions that matter for the observables, and to which degree results obtained under simplifying assumptions can be interpreted quantitatively. We observed that the heuristic of radial polarization pattern implying azimuthal magnetic field structure and azimuthal polarization pattern implying poloidal magnetic fields generally holds for global accretion models, particularly for flows with non-negligible inflow, as expected in RIAFs. The geometric thickness of the disk does not affect the observables significantly in most cases, which might be an argument for the validity of conclusions obtained with a simplifying assumption of the equatorial emission. On the other hand, magnetic field geometry, plasma velocity profile, and BH spin all leave a strong imprint on the resulting images.

While we adjusted parameters to match the observed \m87 flux density of $\sim$0.5\,Jy, our models remained Faraday thin, which is not the case for GRMHD simulations. The observations also indicate an appreciable Faraday depth through partial depolarization \citep{2021ApJ...910L..12E,2021ApJ...910L..13E}. We further inspected this discrepancy by comparing our RIAF model setup to the GRMHD results in the EHT library, discussed by \citet{Dhruv2025}. We found significant differences between the assumed RIAF structure discussed in Sect.~\ref{sec:flow_description} and predictions of GRMHD, such as a much steeper decrease in the gas temperature with radius seen in GRMHD. As a consequence, lower temperatures of the electrons in the emitting region might be a reason for the more pronounced Faraday effects seen in GRMHD \citep{Quataert2000}. Attempts to reproduce the appreciable Faraday depth by roughly matching the RIAF model properties to those of the averaged GRMHD were unsuccessful, however, which led us to the tentative conclusion that either nonaxisymmetric turbulent flow structure or off-equatorial jet base region plasma cause the observed depolarization.

As a consequence of the lack of depolarization, our RIAF models are all very symmetric, resulting in a large amplitude of $\beta_2$ and resolved polarization degree $\langle |m| \rangle$. This is inconsistent with the observations. There is a caveat, related to the fact that we did not explore the effect of the inclination angle. The uncertainties of the observational estimate of the inclination are not smaller than about $5^\circ-10^\circ$, which leaves some space for the fine-tuning. A higher inclination would generally reduce the image symmetry, bringing measurements of $|\beta_2|$ down. For the other remaining observables, most of the RIAF models violate at least some of those. The argument of $\beta_2$ appears to be particularly constraining. One model that fares reasonably well against these constraints is the sub-Keplerian flow with a dipolar magnetic field configuration and a spin of $a_* = 0.5$. This solution visually disagrees with the EHT image reconstruction, however, with a presence of a second ring (see the discussion in Sect.~\ref{sec:4}). More broadly, our results (particularly the argument of $\beta_2$) indicate dominance of poloidal magnetic fields in \m87, and the relatively hgih image brightness asymmetry discourages zero or mildly negative BH spins. It is apparent that the EHT observations have a strong constraining power over the accretion models. The model that is most consistent with the data would necessarily combine poloidal and toroidal magnetic field components, as well as azimuthal and radial velocity patterns.

Our analysis demonstrated that the horizon-scale observables depend in a systematic and physically interpretable way on the magnetic field geometry, plasma velocity structure, and black hole spin. Our main conclusions are summarized below.
\begin{itemize}
    \item The magnetic field geometry (in particular the distinction between the poloidal and toroidal field) strongly affects the structure of the polarized image and the EVPA pattern.
    \item The plasma velocity profile affects multiple image properties, most notably, the asymmetry (which is highest for the Keplerian model), but also the polarimetric observables.
    \item The black hole spin predominantly affects the images by increasing the brightness asymmetry. There are spin-dependent trends in the EVPA structure, but the effects are degenerate with respect to the magnetic field geometry and the velocity profile.
     \item The geometric thickness plays a secondary role under the assumption that the emission is dominated by the accretion disk. 
\end{itemize} 

We discussed the effect of RIAF model parameters on the EHT observables mostly qualitatively. Some of the limitations of this work include fixed assumptions on the power-law structure of the flow and a lack of mixed toroidal-poloidal magnetic field configurations. Fitting a more general RIAF model directly to the EHT data would constitute an interesting follow-up work. The feasibility of a similar approach was recently demonstrated by \citet{Saraer2025}.  
Within the RIAF framework, accretion disk configurations in non-Kerr spacetimes might also be tested without resorting to computationally exhaustive covariant MHD numerical simulations (see \citet{Vincent2021} or the recent developments from \citet{Yin2025}).

\begin{acknowledgements}
We thank George Wong for helpful comments as well as Frederic Vincent and the Paris Observatory in Meudon for hospitality during the work on this project. We thank Jack Livingston for internal refereeing of the manuscript. MW is supported by a~Ramón y Cajal grant RYC2023-042988-I from the Spanish Ministry of Science and Innovation. Saurabh's visit to Paris Observatory was supported by the Procope-Mobilit\"at 2024 grant from the Embassy of France in Germany. This work was supported by the M2FINDERS project funded by the European Research Council (ERC) under the European Union's Horizon 2020 Research and Innovation Programme (Grant Agreement No. 101018682). AT acknowledges the Alexander von Humboldt Foundation for its Fellowship and the Czech Science Foundation (GA{\v C}R) Grant No. 23-07043S. RE warmly acknowledges the generous support of the Institute for Theory and Computation (ITC) at the Center for Astrophysics.
\end{acknowledgements}

\bibliography{ref}{}
\bibliographystyle{aa}
\appendix

\end{document}